\documentclass[%
 reprint,
 superscriptaddress,
 amsmath,amssymb,
 aps,
 prb
]{revtex4-2}

\usepackage{graphicx}
\usepackage{dcolumn}
\usepackage{bm}
\usepackage{url}
\usepackage{multirow}
\usepackage[colorlinks=true,citecolor=blue,urlcolor=blue]{hyperref}
\hypersetup{colorlinks=true, urlcolor=blue, citecolor=cyan, pdfborder={0 0 0}}

\begin{document}


\title{First-principles carrier mobility and optical absorption of strained ZnO with self-consistent Hubbard interactions}
\author{Hong-Guk Min}
\affiliation{Korea Institute for Advanced Study, Seoul 02455, Korea}
\author{Wooil Yang}
\email{Contact Author: yspacefirst@kias.re.kr}
\affiliation{Korea Institute for Advanced Study, Seoul 02455, Korea}
\affiliation{Oden Institute for Computational Engineering and Sciences,
The University of Texas at Austin, Austin, Texas 78712, USA}
\affiliation{Department of Physics, The University of Texas at Austin, Austin, Texas 78712, USA}
\author{Sabyasachi Tiwari}
\affiliation{Oden Institute for Computational Engineering and Sciences,
The University of Texas at Austin, Austin, Texas 78712, USA}
\affiliation{Department of Physics, The University of Texas at Austin, Austin, Texas 78712, USA}
\author{Feliciano Giustino}
\affiliation{Oden Institute for Computational Engineering and Sciences,
The University of Texas at Austin, Austin, Texas 78712, USA}
\affiliation{Department of Physics, The University of Texas at Austin, Austin, Texas 78712, USA}
\author{Young-Woo Son}
\email{Contact Author: hand@kias.re.kr}
\affiliation{Korea Institute for Advanced Study, Seoul 02455, Korea}

\date{\today}

\begin{abstract}
Carrier mobility and optical absorption are key performance parameters of oxide semiconductors in transparent and flexible displays. 
We use a newly developed density-functional perturbation theory with a self-consistent Hubbard correction (DFPT+$U$) to study phonon-limited electron transport and phonon-assisted optical absorption in strained zinc oxide (ZnO). 
This parameter-free approach accounts for electron–phonon interactions and on-site correlation effects simultaneously. Electronic structures and phonon dispersions are computed under three distinct uniaxial strain directions. 
Uniaxial tensile strain up to 4.8\% along [$\bar{1}10$]
is found to increase the room-temperature electron mobility by 19\% while leaving visible-range optical absorption essentially unchanged. 
These results demonstrate that moderate strain can selectively enhance carrier transport without degrading optical transparency, and establish DFPT+$U$ as an effective framework for predicting strain-dependent transport and optical properties in wide-band-gap oxides with implications for strain-engineered display and optoelectronic applications.
\end{abstract}

\maketitle


\section{Introduction}

Semiconductor devices
form the foundation of modern electronics,
with continual improvements in performance and integration driven by advances in materials, interfaces, and scaling~\cite{shockley1949theory,bardeen1948transistor, sze2021physics,moore1965cramming,1050511, bohr200730}. A key example is the metal–  oxide–semiconductor (MOS) architecture, in which thermally grown oxides stabilize the silicon surface and provide a reliable field-effect operation \cite{atalla1959stabilization, kahng1960silicon, hofstein1963silicon, deal1965general}. 
The thin-film transistor (TFT) extended this concept to thin semiconductor films on insulating substrates, forming the basis of large-area electronics \cite{4066878, le1979amorphous, kagan2003thin}. Within this framework, oxide semiconductors, {\it e.g.}, ZnO, NiO, and indium gallium zinc oxide, can serve as TFT channel materials owing to their wide band gaps, optical transparency, and tunable conductivity 
\cite{sato1993transparent, orita2000mechanism, hoffman2003zno, nomura2004room, yabuta2006high, fortunato2012oxide, ohta2004transparent}.

TFTs are widely used as the switching elements in active-matrix display 
backplanes~\cite{brody19736, lechner1971liquid}. The development of flexible electronics has increased demand for mechanically compliant backplanes that maintain high performance under repeated bending \cite{kim2023fabrication, byeon2023recent}. ZnO-based TFTs are promising due to  low temperatures fabrication on polymer substrates,
along with high carrier transport and optical transparency \cite{nomura2004room,fortunato2005fully,park2009flexible,song2010fully, jun2011high,lin2015stable}. During bending, the channel and gate stack experience tensile or compressive strain, affecting carrier mobility and potentially accelerateing bias- and defect-related instabilities \cite{cherenack2010impact,eun2012mechanical}. Although crystalline ZnO thin films are brittle and can develop cracks under strain \cite{sierros2010mechanical,fortunato2002transparent}, solution-processed ZnO TFTs have been reported to remain functional at tensile strains up to 6.35\% \cite{song2010fully}. Strain can also shift the absorption edge via band-gap changes, linking mechanical deformation to optical response \cite{li2007characterization,liu2024asymmetric}. A number of experiments have introduced uniaxial strain into oxide thin films by stretching flexible substrates, bending them during or after film growth, or using piezoelectric actuators \cite{jia2026modulation, shieh2010electromechanical,choi2015prestress,arata2022strain}.
These observations motivate a quantitative microscopic study of strain effects on transport and optical properties of oxide TFT channels.

Carrier mobility and optical absorption in semiconductors are strongly influenced by lattice vibrations through electron–phonon (el-ph) interactions \cite{fan1951temperature,bardeen1950deformation,grimvall1969new}. Density functional theory (DFT) \cite{hohenberg1964, PhysRev.140.A1133}, combined with density functional perturbation theory (DFPT) \cite{baroni1987green,giannozzi1991ab,gonze1995adiabatic,PhysRevA.54.4591}, provides a first-principles framework to compute the lattice dynamics and el-ph interactions \cite{baroni2001phonons, giustino2017electron, jones2015density}. 
Since lattice-related physical properties are closely tied to the electronic structure, 
the well-known errors from the standard (semi)local exchange-correlation functionals~\cite{perdew1981self,perdew1996generalized},
such as underestimated band gaps and over-delocalize electronic states~\cite{mori2008localization,lim2012angle,oba2011point},
can likewise influence the accuracy of lattice-dynamical properties and el-ph interactions.
Among several methods to overcome these errors, 
the DFT+$U$ \cite{anisimov1997first, dudarev1998electron, himmetoglu2014hubbard} approach is often adopted with the Hubbard parameter $U$ chosen empirically or determined self-consistently \cite{agapito2015reformulation, cococcioni2005linear, kulik2006density,mosey2007ab,mosey2008rotationally}. Building on this idea, DFPT+$U$ \cite{floris2011vibrational, floris2020hubbard, yang2025first} extends DFPT to enable direct evaluation of phonons and el-ph interactions from a DFT+$U$ ground state. 
In ZnO, Hubbard corrections to the Zn $3d$ manifold ($U_d$) and/or the O $2p$ manifold ($U_p$) significantly improve the band gap \cite{lany2008assessment, lim2012angle,gopal2015improved}. Notably, a self-consistent $U_p$ alone can reproduce experimental phonon-limited mobility and optical absorption \cite{yang2025first}.

In this work, we investigate phonon-limited electron transport in ZnO under uniaxial tensile strain. 
The direct optical absorption as well as phonon-assisted indirect one are also computed with varying strains. 
These physical quantities are obtained within the DFPT+$U$ framework, in which $U$ is computed self-consistently and the el-ph matrix elements are evaluated from a DFT+$U$ ground state.
According to a recent work \cite{yang2025first}, we adopt a minimal Hubbard correction applied only to the O $2p$ orbital for reliable band structures and el-ph interactions. We find that a uniaxial tensile strain of $4.8\%$ enhances the room-temperature electron mobility along the strain direction by up to $19\%$ relative to the unstrained case. Over the same strain range, the optical absorption spectrum shows no appreciable change below $3.2\,\mathrm{eV}$, indicating that visible-range transparency is preserved under strain. 
Our work shows that the newly developed DFPT+$U$ framework with self-consistent Hubbard corrections enables quantitative understanding of complex physical processes in technologically important semiconducting materials
without adjustable parameters. 


\section{Computational methods}
First-principles calculations were performed using our in-house modification~\cite{lee2020first, yang2021ab, inhouse} of the \textsc{quantum espresso} package \cite{giannozzi2009quantum}. Scalar relativistic norm-conserving pseudopotentials were used \cite{hamann2013optimized}, and the local density approximation (LDA) \cite{perdew1981self} was employed for the exchange-correlation functional. The plane-wave basis set was constructed with a cutoff of $110$ Ry. Structural relaxation and self-consistent charge density calculations were carried out on an $18\times18\times18$ Monkhorst-Pack k-point grid \cite{monkhorst1976special} in the first Brillouin zone (BZ). 
The self-consistent Hubbard parameter for each strain condition is obtained based on the ACBN0 approach \cite{agapito2015reformulation,lee2020first, yang2021ab}. 
The L\"owdin orthonormalized atomic orbital (LOAO) is used as projectors to obtain $U_p$~\cite{lee2020first, yang2021ab}. 

Strained ZnO structures were generated from the relaxed pristine structure by stretching the lattice constant along the strain direction. A rectangular unit cell was adopted, and the lattice vectors orthogonal to the strain direction were relaxed until the Hellmann-Feynman forces were below $2\times 10^{-6}$Ry/Bohr. 
The optimized lattice parameters and the corresponding Hubbard parameters for the strained structures are summarized in Table~\ref{tab:y_strained_cell}. Phonons were computed using DFPT+$U$ on a $7\times7\times7$ $q$-point grid in the BZ. 

Electron-phonon matrix elements, transport, and optical properties were computed using the EPW code \cite{ponce2016epw, lee2023electron, giustino2007electron} with the \textsc{wannier90} package \cite{pizzi2020wannier90} in library mode. Hybrid orbitals were used as the initial Wannier projectors, and a frozen energy window of $\sim15$ eV centered around the Fermi level was adopted. The frozen-window parameters for each strain condition are summarized in Table~\ref{tab:wan}. Mobility and optical absorption were evaluated using the Wannier-Fourier interpolation \cite{giustino2007electron, mostofi2014updated} from the same coarse meshes of $14\times14\times14$ $k$-point and $7\times7\times7$ $q$-point onto dense grids tailored to each observable. For mobility calculations, interpolation was performed onto $144\times144\times144$ grid for both $k$ and $q$ points. In this work, we calculate the mobility of electrons near the conduction-band minimum; hole mobility is not included. The electron carrier concentration was set to $10^{16}~\mathrm{cm}^{-3}$. The mobility was calculated using the iterative solution of the linearized Boltzmann transport equation \cite{ponce2018towards}. For optical properties, the interpolation used a $70\times70\times70$ $k$-point grid together with a $7\times7\times7$ $q$-point grid. To accurately capture low-energy phonon modes that can dominate el-ph scattering \cite{lihm2024self}, the acoustic phonon energy cutoff was set to $0.01$ meV, which is lower than commonly used values. The absorption spectrum including both direct and phonon-assisted transitions was calculated using quasi-degenerate perturbation theory \cite{tiwari2024unified}, with a energy window parameter of 10~meV.

\section{Results}
\subsection{Atomic structure of uniaxially strained ZnO from DFT+U}

\begin{table}[t]
\caption{\label{tab:y_strained_cell}Relaxed lattice parameters and self-consistent Hubbard $U_p$ values for $\varepsilon_{yy}$-strained ZnO. Here $a'$ and $c'$ are the in-plane and out-of-plane lattice constants, respectively, and $l$ is the in-plane distortion parameter defined in the main text.}
\begin{ruledtabular}
\centering
\begin{tabular}{c|cccc}
$\varepsilon_{yy}$ Strain (\%) & $a'$ (\text{\AA}) & $l$ & $c'$ (\text{\AA}) & $U_p$ (eV) \\
\hline
0  & 3.18 & $\sqrt{3}$ & 5.10 & 2.8810 \\
0.8  & 3.17 & 1.75      &  5.09 & 2.8818 \\
1.8  & 3.15 & 1.78      &  5.07 & 2.8829 \\
2.8  & 3.14 & 1.80      &  5.06 & 2.8845 \\
3.8  & 3.13 & 1.83      &  5.04 & 2.8869 \\
4.8  & 3.11 & 1.85      &  5.02 & 2.8906 \\
\end{tabular}
\end{ruledtabular}
\end{table}

\begin{figure}[b] 
\includegraphics[width=0.45\textwidth]{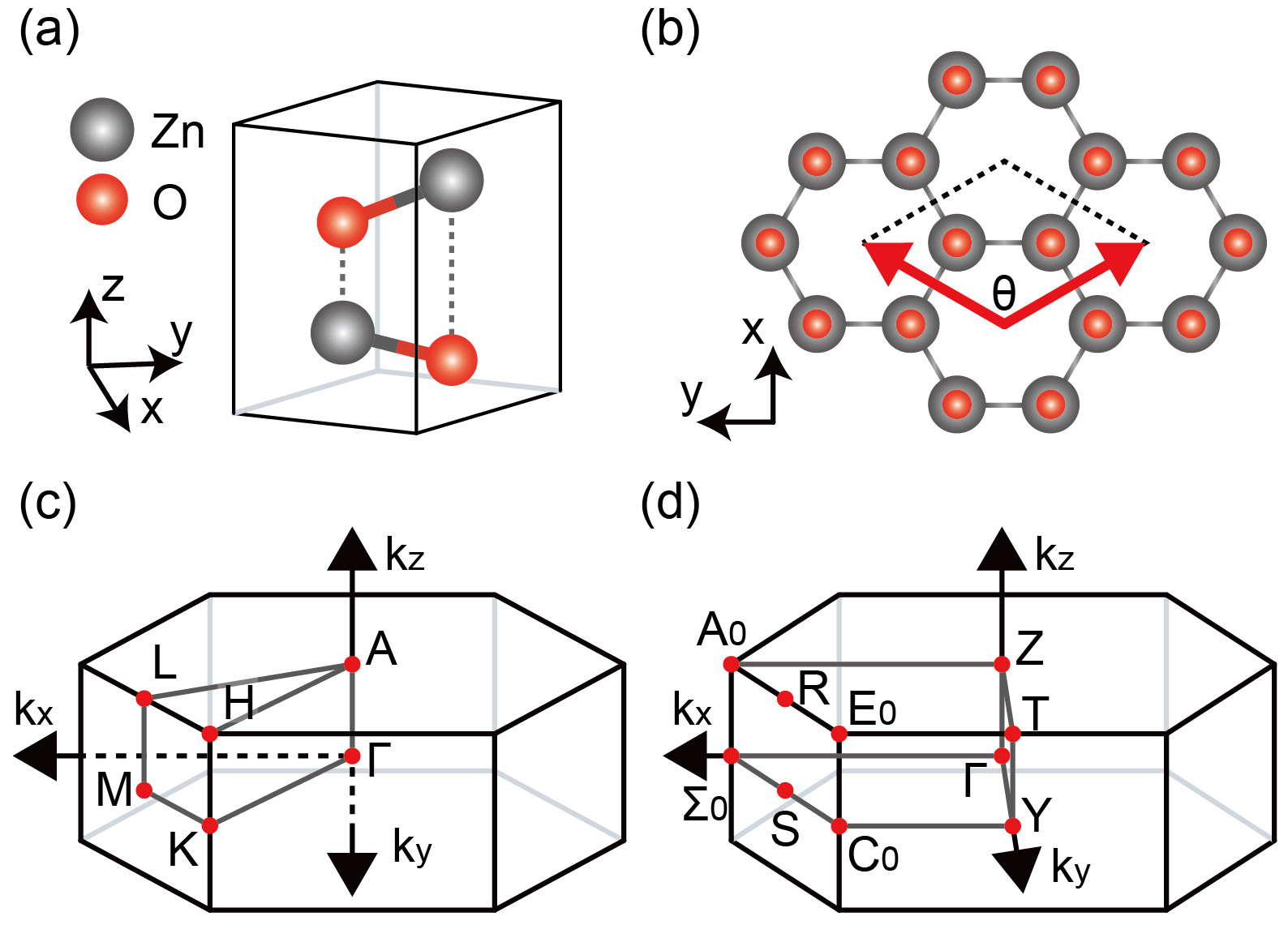}
\caption{\label{fig:atom}
Atomic structure of ZnO and the corresponding Brillouin zones (BZs). (a) Primitive unit cell of unstrained wurzite ZnO. (b) Top view of the atomic structure. Red arrows indicate the primitive lattice vectors, while dashed line shows the boundary of the unit cell. (c) Hexagonal and (d) orthorhombic BZs, with high-symmetry points highlighted with red dots.
}
\end{figure}

We begin by describing the atomic structure of pristine and strained ZnO. 
Figure~\ref{fig:atom} (a) shows the primitive unit cell of unstrained ZnO in the hexagonal space group $P6_3mc$ (No.~186). The Zn and O atoms connected by the dotted line share the same in-plane coordinates and differ only in their $z$ positions. A top view is shown in Fig.~\ref{fig:atom} (b), where the red arrows indicate the in-plane primitive lattice vectors and the dashed line marks the unit cell boundary. The lattice vectors are chosen in the conventional way for a hexagonal cell, 
$\bm a_1 = \bigl(\tfrac{1}{2}a, \tfrac{\sqrt{3}}{2}a, 0\bigr)$, 
$\bm a_2 = \bigl(\tfrac{1}{2}a, -\tfrac{\sqrt{3}}{2}a, 0\bigr)$, and 
$\bm a_3 = (0, 0, c)$.
We fully relax the atomic structure within the DFT+$U$ framework, where the on-site Hubbard parameter for the O $2p$ orbitals, $U_p$, is determined self-consistently for each updated geometry. The optimized in-plane and out-of-plane lattice parameters are $a = 3.18$~\AA{} and $c = 5.10$~\AA{}, respectively, with $U_p = 2.88$~eV. These values are consistent with previously reported structural parameters for wurtzite ZnO, within $\pm 0.1$~\AA{} \cite{abrahams1969remeasurement,schulz1979structure, kisi1989u}. 

Starting from the relaxed pristine structure, we consider uniaxial tensile strain along three representative crystallographic directions: the zigzag $[110]$, armchair $[\bar{1}10]$, and out-of-plane $[001]$ directions. 
These are denoted as $\varepsilon_{xx}$, $\varepsilon_{yy}$ and $\varepsilon_{zz}$, respectively. 
Owing to the rotational symmetry about the $[001]$ axis, the basal plane of pristine wurtzite ZnO is isotropic for second-rank linear-response properties~\cite{nye1985physical}. 
An in-plane uniaxial strain lowers this parent hexagonal symmetry into an orthorhombic cell by selecting a specific in-plane axis. 
Therefore, the armchair- and zigzag-strained structures can be regarded as different orientation variants of the same symmetry-breaking distortion. 
In contrast, the $\varepsilon_{zz}$ strain preserves the symmetry of pristine ZnO and changes only the lattice parameters and internal atomic coordinates. 
Since in-plane tensile strain is most relevant for thin-film applications of ZnO, we focus on effects of $\varepsilon_{yy}$ strain in this work. 
The atomic and electronic structures for $\varepsilon_{xx}$ 
and $\varepsilon_{zz}$ strains are provided in Appendix D for comparison.

Upon application of $\varepsilon_{yy}$ strain, the parent hexagonal symmetry of $P6_3mc$ is lowered to the orthorhombic space group $Cmc2_1$ (No.\,36). The deformation can be parameterized by the angle $\theta$ between the in-plane primitive lattice vectors, as illustrated in Fig.~\ref{fig:atom} (b). We define the strained lattice vectors as
$
\bm a_1' = \Bigl(\tfrac{1}{2}a', \tfrac{l}{2}a', 0\Bigr), 
\bm a_2' = \Bigl(\tfrac{1}{2}a', -\tfrac{l}{2}a', 0\Bigr),  \bm a_3' = (0, 0, c')$,
where $a'$ and $c'$ are the strained cell parameters and $l$ controls the in-plane distortion. The pristine hexagonal cell is recovered when $l = \sqrt{3}$, whereas an $n\%$ tensile strain corresponds to
$l a' = (1+n/100)\times \sqrt{3} a$.
The angle between the in-plane primitive vectors then becomes $\theta = 2 \tan^{-1}(l)$.
Under uniaxial strain, the Poisson effect governs the change in the lattice perpendicular to the strain direction. In our structural optimization, we fix the $y$ components of the in-plane lattice vectors, $\pm (l/2)a'$, which set the target strain value along the $[\bar{1}10]$ direction. The lattice constants along the $x$ and $z$ directions are fully relaxed to obtain stable strained structures. For each strain condition, the Hubbard parameter $U_p$ is recomputed self-consistently within the ACBN0 scheme. The relaxed lattice parameters and the corresponding $U_p$ values are summarized in Table~\ref{tab:y_strained_cell}. We find that $U_p$ is robust against strain; even at the maximum tensile strain considered, $U_p = 2.89$~eV, differing by only $0.33\%$ from the pristine value. Fig.~\ref{fig:atom} (c) and (d) show the first Brillouin zones (BZs) and high symmetry points for the hexagonal and orthorhombic  lattices, respectively.

\subsection{Electron and phonon energy dispersions}
\begin{figure}[t] 
\includegraphics[width=0.45\textwidth]{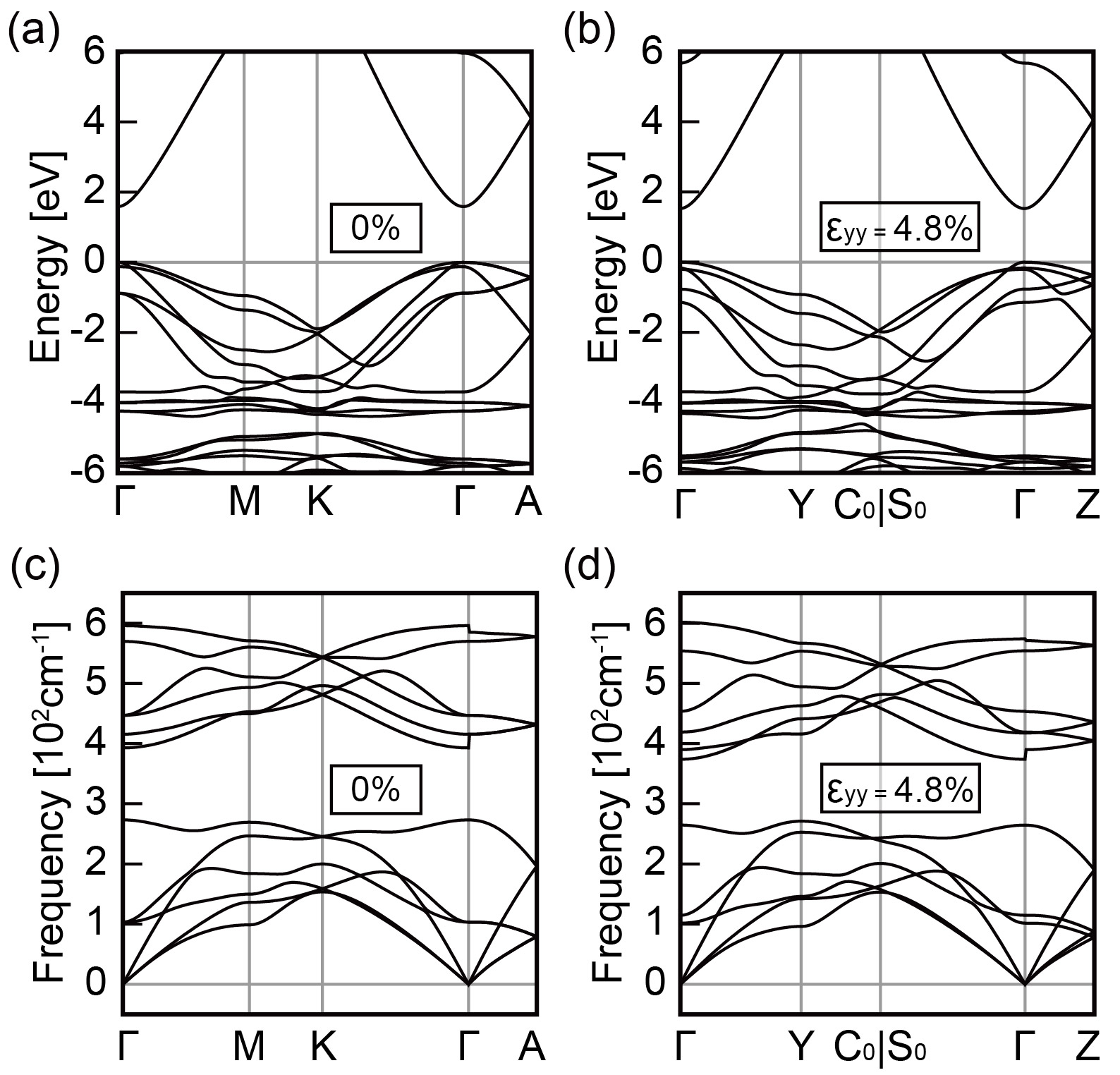}
\caption{\label{fig:elph}
Electron and phonon band structures of ZnO under strain. (a), (b) Electronic structure of pristine and 4.8\% $\varepsilon_{yy}$-strained ZnO, respectively. (c), (d) Corresponding phonon dispersions for pristine and 4.8\% $\varepsilon_{yy}$-strained ZnO. 
}
\end{figure}

DFT calculations employing (semi)local approximations for exchange-correlation functionals~\cite{perdew1981self,perdew1996generalized} typically compute ZnO band gaps in the range of $0.7\sim0.9$~eV \cite{kohan2000first, janotti2009fundamentals}, which are quite smaller than the experimental value of $3.3\sim3.4$~eV \cite{mang1995band, dong2004electronic, reynolds1999valence}. This discrepancy directly affects the accuracy of computed el-ph couplings and related phonon-limited properties. The inclusion of $U_p$ is known to partially restore the band gap and improve agreement with experimental phonon-limited behavior \cite{gopal2015improved, yang2025first}. Figure~\ref{fig:elph} (a) shows the electronic band structure of unstrained ZnO obtained from our DFT+$U$ calculations along the high-symmetry path in Fig.~\ref{fig:atom} (c). For the relaxed pristine structure, the calculated band gap is $1.59$~eV, which is significantly larger than the semilocal DFT value, although it remains smaller than the experimental gap. The phonon dispersion of unstrained ZnO obtained from DFPT+$U$ is shown in Fig.~\ref{fig:elph} (c).

For comparison, the band structure under $4.8\%$ $\varepsilon_{yy}$ strain is shown in Fig.~\ref{fig:elph} (b), and the corresponding phonon dispersion is shown in Fig.~\ref{fig:elph} (d). The strained system remains insulating with a direct gap near $\Gamma$ and shows no abrupt band reconstruction. Along the $\Gamma$–Z direction, the reduction of symmetry from hexagonal to orthorhombic lifts certain degeneracies and leads to additional splittings in both the electronic and phonon dispersions, while leaving the overall electronic bandwidths nearly unchanged. These changes in the energy landscape modify the energy-matching conditions for scattering processes relevant to mobility and optical absorption. The band gap decreases linearly with tensile strain at a rate of $\sim 12$~meV per $1\%$ strain, comparable to the experimentally reported value of $\sim 16$~meV per $1\%$ tensile strain for bent ZnO thin films \cite{liu2024asymmetric}. No imaginary phonon frequencies are found anywhere in the BZ, confirming the dynamical stability of the strained structures. The full electronic and phonon band structures for all $\varepsilon_{yy}$ strain conditions are provided in Appendix A.

\subsection{Strain-dependent mobility and optical properties}

\begin{table}[t]
\caption{\label{tab:mobility}Temperature-dependent electron mobility of pristine ZnO.}
\begin{ruledtabular}
\centering
\begin{tabular}{c|ccc}
\multirow{2}{*}{Temperature (K)} & \multicolumn{3}{c}{Mobility ($\mathrm{cm}^2/\mathrm{V} \cdot \mathrm{s}$)} \\ & $\mu_x^0$ & $\mu_y^0$ & $\mu_z^0$ \\
\noalign{\vskip 2pt}
\hline
\noalign{\vskip 2pt}
50  & 878.18 & 878.18 & 873.29 \\
100  & 657.11 & 657.11 & 588.20 \\
150  & 490.48 & 490.48 & 429.24 \\
200  & 345.43 & 345.43 & 302.63 \\
250  & 242.47 & 242.47 & 214.00 \\
300  & 176.36 & 176.36 & 156.99 \\
350  & 134.30 & 134.30 & 120.56 \\
400  & 106.74 & 106.74 & 96.57 \\
\end{tabular}
\end{ruledtabular}
\end{table}

We interpolate the el-ph matrix elements obtained from DFPT+$U$ using the EPW package \cite{ponce2016epw, lee2023electron, giustino2007electron, yang2025first}. 
Reliable interpolation requires a Wannier Hamiltonian that faithfully reproduces the underlying DFT+$U$ electronic structure. Our Wannier Hamiltonian accurately reproduces the DFT+$U$ band dispersion, particularly near the conduction-band edge that governs phonon-limited mobility and the near-edge optical response. The Wannier-interpolated band structure and Wannierization parameters are reported in Appendix B. Figures~\ref{fig:mob} (a)-(c) show the temperature-dependent electron mobility along the $x$, $y$, and $z$ directions, respectively, under $\varepsilon_{yy}$ strain. For each temperature, the strained mobilities are normalized by the pristine values summarized in Tab.\,\ref{tab:mobility}, so the plotted quantity represents the relative mobility change in percent upon applications of the strains. The strained cases are shown in shades of blue, with increasing strain indicated by progressively darker curves. Our results show that $\varepsilon_{yy}$ strain enhances both in-plane mobility components. Along the strain direction, the mobility $\mu_y$ at $300$~K increases by $19\%$ for the maximum strain of $4.8\%$, while the perpendicular in-plane component $\mu_x$ increases by $4.2\%$. In contrast, the out-of-plane mobility $\mu_z$ decreases by $8\%$ at room temperature for the maximum strain. 

\begin{figure}[t] 
\includegraphics[width=0.47\textwidth]{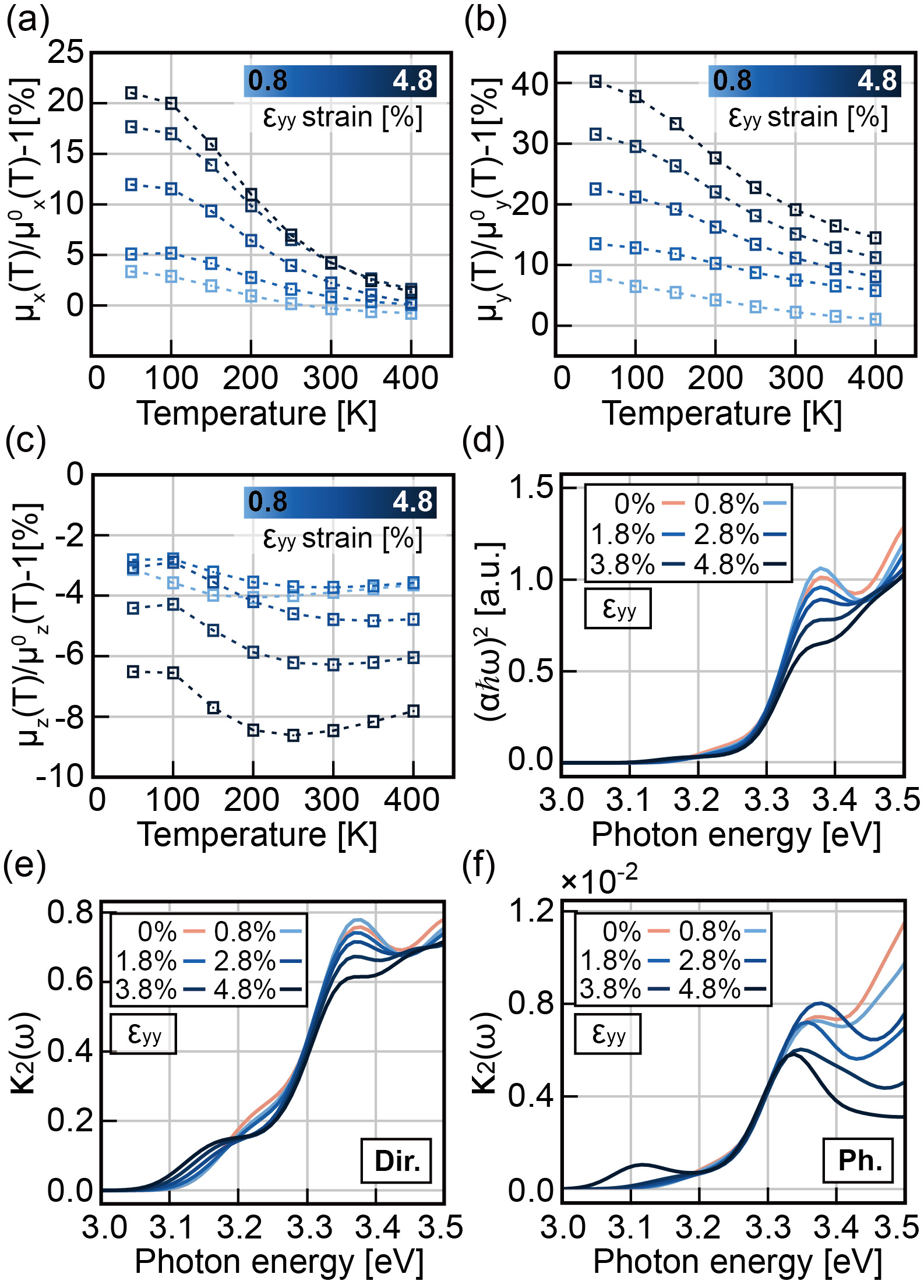}
\caption{\label{fig:mob}
Transport and optical properties of ZnO under strain. (a)–(c) Relative change of the phonon-limited electron mobility along the (a) $x$, (b) $y$ (strain), and (c) $z$ directions. The changes in percent are shown with respect to the $\mu_i^{0}(\mathrm{T})$, which denotes the pristine mobility at each temperature. The strained mobilities are colored in shades of blue, with darker colors indicating larger $\varepsilon_{yy}$ strain ranging from $0.8\%$ to $4.8\%$. (d) Optical absorption spectra as a function of photon energy. The pristine spectrum is shown in pink, while the $\varepsilon_{yy}$-strained spectra are plotted in progressively darker blue, consistent with the color scheme used for the mobility. (e),(f) Imaginary dielectric function of $\kappa_2$ for (e) direct and (f) phonon-assisted indirect transitions. Note that y-axis scale in panel (f) is 100 times smaller than that in panel (e).
}
\end{figure}

The phonon-assisted optical absorption spectra for different strain values are shown in Fig.~\ref{fig:mob} (d). Since DFT+$U$ still underestimates the electronic band gap, we apply a scissor correction to align the calculated gap with experiment. Following Ref.~\cite{zacharias2016one}, a scissor operator $(E_c - E_v)/(E_c - E_v + \Delta)$ is introduced, where $E_c - E_v$ is the DFT+$U$ gap of the pristine structure ($E_g^{\text{DFT}+U} = 1.59$~eV) and $\Delta$ is the discrepancy with experiment, $\Delta=E_g^{\text{exp}} - E_g^{\text{DFT}+U} = 1.61$~eV. We set $E_g^{\text{exp}}=3.20$~eV as a conservative reference gap for assessing visible-light transparency. While the scissor correction rigidly shifts the optical absorption spectrum along the photon-energy axis, the overall trends with strain remain largely unchanged. The electronic band gap decreases by $\sim 12$~meV per $1\%$ strain, a small change compared with $\Delta=1.61$~eV. Thus, for simplicity and to enable direct comparison across strain conditions, the same scissor correction is applied to all strained structures. After this correction, a pronounced peak appears around a photon energy of $3.38$ eV. Its intensity increases at $0.8\%$ strain and then gradually decreases as the strain is further increased. In the visible range ($1.6-3.3$ eV), the absorption spectrum remains essentially unchanged, whereas above $3.3$~eV the absorption decreases systematically with increasing tensile strain.

To further examine the role of electron-phonon scatterings in the optical absorption, we calculate the imaginary part of the dielectric function, $\kappa_2(\omega)$ with varying strain. 
We note that the optical absorption $\alpha$ in Fig.~\ref{fig:mob} (d) is obtained by 
$\alpha(\omega)=
\frac{
\omega \kappa_2(\omega)
}
{
cn(\omega)
}
$
where $\omega$ is the photon frequency, $c$ the speed of light, and 
$n^2(\omega)=
[
\sqrt{\kappa^2_1(\omega)+
    \kappa^2_2(\omega)
        }
       +\kappa_1(\omega)
]/2
$.
Here, $\kappa_1$ is a real part of dielectric function. 
Figures~\ref{fig:mob} (e) and (f) show the contributions to $\kappa_2$ from direct and phonon-assisted indirect transitions, respectively. The direct contribution is much larger than the phonon-assisted contribution, indicating that direct transitions dominate the optical response. The direct transition feature at photon energies of $3.1–3.2$ eV reflects the reduction of the band gap under strain, while the plateau-like structure near $3.2$ eV arises from strain-induced modifications in the electronic energy dispersion near the valence and conduction band edges. For both direct and phonon-assisted contributions, the main peak near $3.38$ eV is gradually suppressed as strain increases. For the phonon-assisted contribution, a weak peak develops in the visible range with increasing strain. However, consistent with the absorption spectra in Fig.~\ref{fig:mob} (d), its magnitude is too small to produce any significant change in visible-light absorption. Overall, these results indicate that ZnO thin films remain optically transparent under $\varepsilon_{yy}$ strain, supporting their use as channel materials in transparent semiconductor devices.

\subsection{Discussion on the origin of mobility enhancement}
\begin{figure}[t] 
\includegraphics[width=0.48\textwidth]{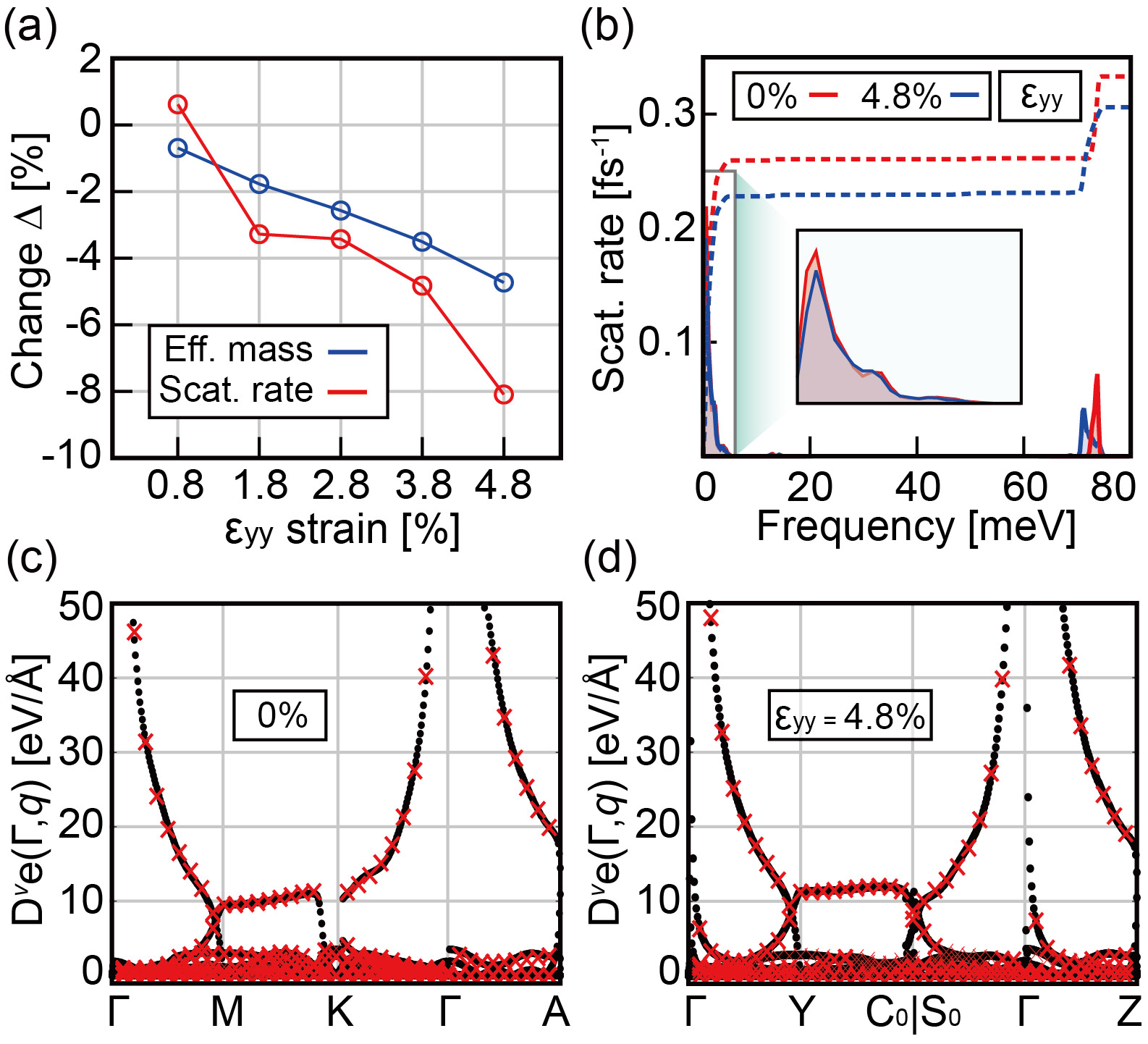}
\caption{\label{fig:effm}
Strain dependence of el-ph interactions. (a) Relative change in the effective mass along the strain direction and the el–ph scattering rate as a function of applied $\varepsilon_{yy}$ strain, normalized to the pristine values. Blue and red circles denote effective mass and scattering rate, respectively. (b) Spectral decomposition of the scattering rate. Solid red (blue) curves show the scattering-rate contributions for the 0\% (4.8\%) strained case. Dotted curves of the corresponding colors indicate the cumulative scattering rates. Inset: Magnified view for phonon energies below $6$~meV. (c), (d) Deformation potential Eq.~\ref{Eq:deformation} of the lowest conduction band for the (c) 0\% and (d) 4.8\% $\varepsilon_{yy}$-strained cases. The black dots are obtained from Wannier interpolation and the red crosses are from direct DFPT+$U$. 
}
\end{figure}

In this section, we discuss the microscopic origin of the mobility enhancement under uniaxial strain. We first examine the electron effective mass at the conduction-band minimum (CBM) along the strain direction. Near the band edge, the lowest conduction band is well described by a quadratic dispersion within an energy window of $80$~meV above the CBM, allowing the effective mass to be extracted from the inverse band curvature. For pristine ZnO, the electron effective mass along the strain direction is obtained as $m^\ast = 0.27\,m_e$, where $m_e$ is the free-electron mass. As shown in Fig.~\ref{fig:effm} (a), this value decreases by $4.73\%$ at the maximum $\varepsilon_{yy}$ strain, indicating enhanced conduction band dispersion. 

Previous studies of wurtzite ZnO have shown that the lowest conduction band and the topmost valence band can be described by a tight-binding model composed of Zn $4s$ and O $2p$ orbitals \cite{kobayashi1983semiempirical,ivanov1981electronic}. In a simplified picture based on an $s$-$p$ bond chain under tensile strain, the increased lattice periodicity along the strain direction enhances the band curvature, while the accompanying increase in the bond length is expected to reduce the hopping amplitude and flatten the bands. However, the directional bandwidth of the conduction band does not decrease under $\epsilon_{yy}$ strain, suggesting that the effective hopping is not suppressed. This behavior can be understood from the directional dependence of the hopping amplitude in the Slater--Koster form \cite{slater1954simplified}. As shown in Fig.~\ref{fig:A4} (b), the tilted Zn--O $sp$-$\sigma$ bond lies in the $yz$ plane. Thus, the hopping contribution along the $y$ direction is proportional to $(d_y/r_1)t_{sp\sigma}$, where $t_{sp\sigma}$ is the bond integral, $r_1$ is the bond length, and $d_y$ is the projected Zn--O distance in the strain direction. Although tensile strain elongates $r_1$, it also increases $d_y$, compensating for the reduction of the radial hopping amplitude. In addition, the band gap decreases under $\varepsilon_{yy}$ strain, further enhancing the band-edge curvature through stronger interband coupling. The full $\varepsilon_{yy}$ strain dependence of the effective mass along all three principal axes is provided in Appendix C.

The change in electron effective mass alone is insufficient to account for the sizable enhancement of the in-plane mobility, which reaches $19\%$ at $300$~K. To quantify the role of electron-phonon scattering, we evaluate the partial scattering rate at $300$~K from the electron relaxation time, defined as
$
 {\tau^{-1}_{n \bm k \rightarrow m\bm k+\bm q}}
= \frac{2\pi}{\hbar}
\sum_{\nu}
| g_{mn\nu}(\bm k,\bm q) |^{2}
[
(n_{\bm q \nu} + 1 - f^{0}_{m \bm k+\bm q})
\delta (\epsilon_{n \bm k} - \epsilon_{m \bm k+\bm q} - \hbar\omega_{\bm q \nu})
 + (n_{\bm q \nu} + f^{0}_{m,\bm k+\bm q})
\delta(\epsilon_{n\bm k} - \epsilon_{m,\bm k+\bm q} + \hbar\omega_{\bm q \nu})
]$
~\cite{ponce2020first}.
Here, $g_{mn\nu}(\bm{k},\bm{q})$  is the el-ph matrix element, $n_{\bm q \nu}$ is the Bose-Einstein phonon distribution, and $f^0_{n,\bm k}$ is the equilibrium electron distribution function. The quantities $\epsilon_{n \bm k}$ and $\hbar \omega_{\bm q \nu}$ denote the electron and phonon energies, respectively, while $n(m)$ and $\nu$ label electronic band indices and phonon branches, respectively. To obtain a representative total scattering rate relevant for transport, we sum over electronic states with energy $\tfrac{3}{2}k_\mathrm{B}T$ above the CBM, which captures the dominant contribution to the mobility \cite{ponce2019origin}. The resulting relative change in the total scattering rate is shown in Fig.\,\ref{fig:effm} (a) as a red solid line. At the maximum strain, the total scattering rate decreases by approximately $8\%$ relative to the pristine case. 

To further analyze the phonon contribution, we decompose the scattering rate as a function of phonon frequency \cite{ponce2021first}, as illustrated in Fig.\,\ref{fig:effm} (b). The dominant contribution originates from acoustic modes below $5$~meV, while a smaller contribution appears from optical modes around $70$~meV. Comparing the $0\%$ and $4.8\%$ cases, we observe a clear reduction in the amplitude of both the acoustic and optical peaks. The dashed curves, which show the cumulative scattering rate up to a given phonon frequency, indicate that the overall decrease is driven primarily by the reduced acoustic contribution. Although the optical peak height is also suppressed, its increased width compensates for the reduction in amplitude. To assess whether strain directly weakens the el-ph coupling strength, we examine the strain dependence of the el-ph matrix element $ g_{mn\nu}(\bm{k},\bm{q})$ via the deformation potential $D^\nu(\Gamma,q)$ that can be written as 
\begin{align}
D^\nu(\Gamma,q) = \frac{1}{N}\bigg[ 2 \rho V \omega_{\bm q \nu} \sum_{nm} \bigl| g_{mn\nu}(\bm{k},\bm{q}) \bigr|^{2}\bigg]^\frac{1}{2}.
\label{Eq:deformation}
\end{align}
Here, $N$ is the number of states, $\rho$ is the mass density of crystal, and $V$ is the unit cell volume. 
We note that Eq.~\ref{Eq:deformation} provides a phase-independent measure of the el-ph coupling near $\Gamma$ \cite{ponce2021first, yang2025first}.

Figures~\ref{fig:effm} (c) and (d) show the deformation potential for unstrained and 4.8\% $\varepsilon_{yy}$-strained ZnO, respectively. The black dots are obtained from interpolation, while the red crosses correspond to the direct DFPT+$U$ evaluation. The close agreement between the two confirms the fidelity of the Wannier interpolation. The total deformation potential shows no appreciable strain-induced reduction of the el-ph coupling near $\Gamma$, indicating that a uniform suppression of matrix elements is not the primary origin of the reduced scattering rate. Instead, the decrease in scattering is more naturally attributed to
strain-induced changes in the electronic dispersion and phonon spectrum,
which modify the scattering phase space and energy-matching conditions. A Drude-like analysis of the mobility shows that the effective mass affects not only the carrier velocity but also the scattering rate through the scattering density of states~\cite{83vn-7zyw}. In the limiting case of a anisotropic three-dimensional parabolic band with dominant acoustic-phonon scattering and nearly unchanged el-ph matrix elements, the mobility along the strain
direction scales as $\mu_y \propto [m_y\sqrt{m_xm_ym_z}]^{-1}$, where
$m_i$ is the band effective mass along direction $i$.
Applying this relation to the effective-mass changes along all three directions gives a 7\% increase in mobility. This estimate is larger than the roughly 5\% increase expected from $m_y^{-1}$ alone, but remains well below the calculated 19\% enhancement of $\mu_y$ at 300~K. Thus, changes in the effective mass contribute to the mobility enhancement, while other factors affecting the scattering energy--matching conditions are also important.

\section{Conclusion}
We have investigated phonon-limited electron transport and optical absorption in wurtzite ZnO under uniaxial tensile strains. Electron–phonon interactions were evaluated within the DFPT+$U$ framework, with the Hubbard parameter applied to the O $2p$ orbitals and determined self-consistently for each strained structure. Tensile strain enhances the in-plane electron mobility at room temperature, while maintaining low visible-range optical absorption. Microscopic analysis indicates that the mobility enhancement is driven primarily by a reduction in acoustic-phonon scattering, with an additional contribution from the moderate decrease of the conduction-band effective mass along the strain direction. These results identify uniaxial strain as an effective route to tune the transport properties of ZnO, offering quantitative guidance for the optimization of oxide-semiconductor TFT channels in flexible and transparent display backplanes. Taken together, our work highlights the value of consistent parameter-free first-principles treatments of electron–phonon coupling with Hubbard interactions in strained semiconductors when assessing performance-relevant metrics such as mobility and optical transparency.

\begin{acknowledgments}
H.-G. M., W.Y. and Y.-W. S. were supported by KIAS individual Grants (No. CG101901, No. QP090102 and No. CG031509). 
Computations were supported by the CAC of KIAS.
Y.-W. S. was also supported by the National Research Foundation of Korea grant funded by the Korea government (MSIT) (RS-2026-25508298). 
This research was partly supported by SUPREME, one of seven centers in JUMP 2.0, a Semiconductor Research Corporation (SRC) program sponsored by DARPA (Mobility calculations in oxides); by the Computational Materials Science program of U.S. Department of Energy, Office of Science, Basic Energy Sciences under Award DE-SC0020129 (DFPT+U development in EPW); and by the U.S. National Science Foundation, DMREF Grant No. 2119555 (Calculations of optical spectra with QDPT). The authors acknowledge the Texas Advanced Computing Center (TACC) at The University of Texas at Austin for providing HPC resources, including the Frontera and Lonestar6 systems, that have contributed to the research results reported within this paper. URL: http://www.tacc.utexas.edu. This research used resources of the National Energy Research Scientific Computing Center, a DOE Office of Science User Facility supported by the Office of Science of the U.S. Department of Energy under Contract No. DE-AC02-05CH11231. This research used resources of the Argonne Leadership Computing Facility, which is a DOE Office of Science User Facility supported under Contract DE-AC02-06CH11357.
\end{acknowledgments}

\vspace{1em}

\appendix

\setcounter{figure}{0}
\renewcommand{\thefigure}{A\arabic{figure}}
\setcounter{table}{0}
\renewcommand{\thetable}{A\arabic{table}}

\begin{figure}[b]
\centering
\includegraphics[width=0.48\textwidth]{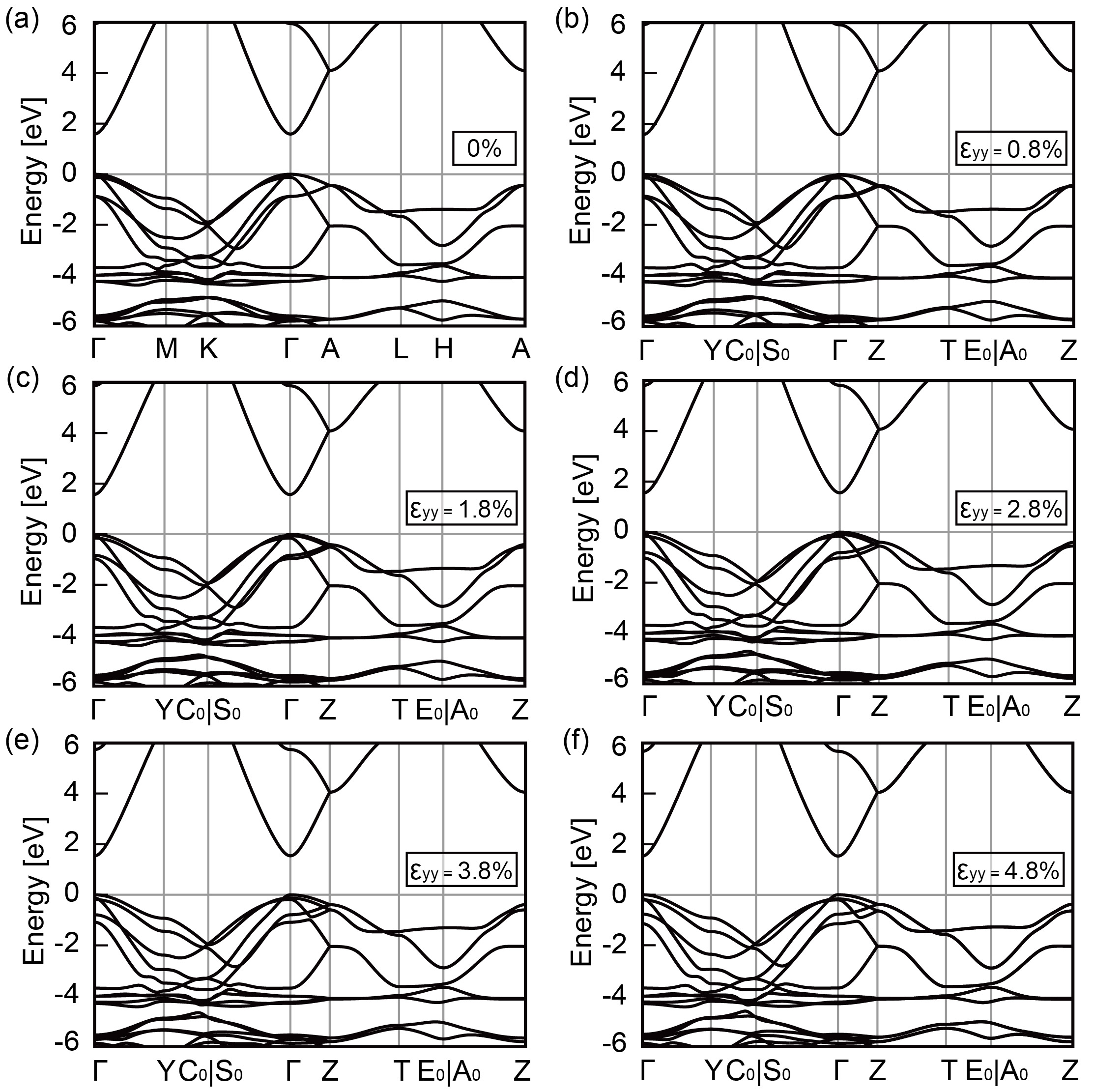}
\caption{\label{fig:A1}
Electronic band structures of ZnO along high-symmetry lines. (a) Pristine ZnO in the hexagonal Brillouin zone. (b)--(f) $\varepsilon_{yy}$-strained ZnO in the orthorhombic Brillouin zone; the applied strain is indicated in each panel.
}
\end{figure}

\section{Electron and phonon band structure for strained ZnO}

\begin{figure}[t]
\centering
\includegraphics[width=0.48\textwidth]{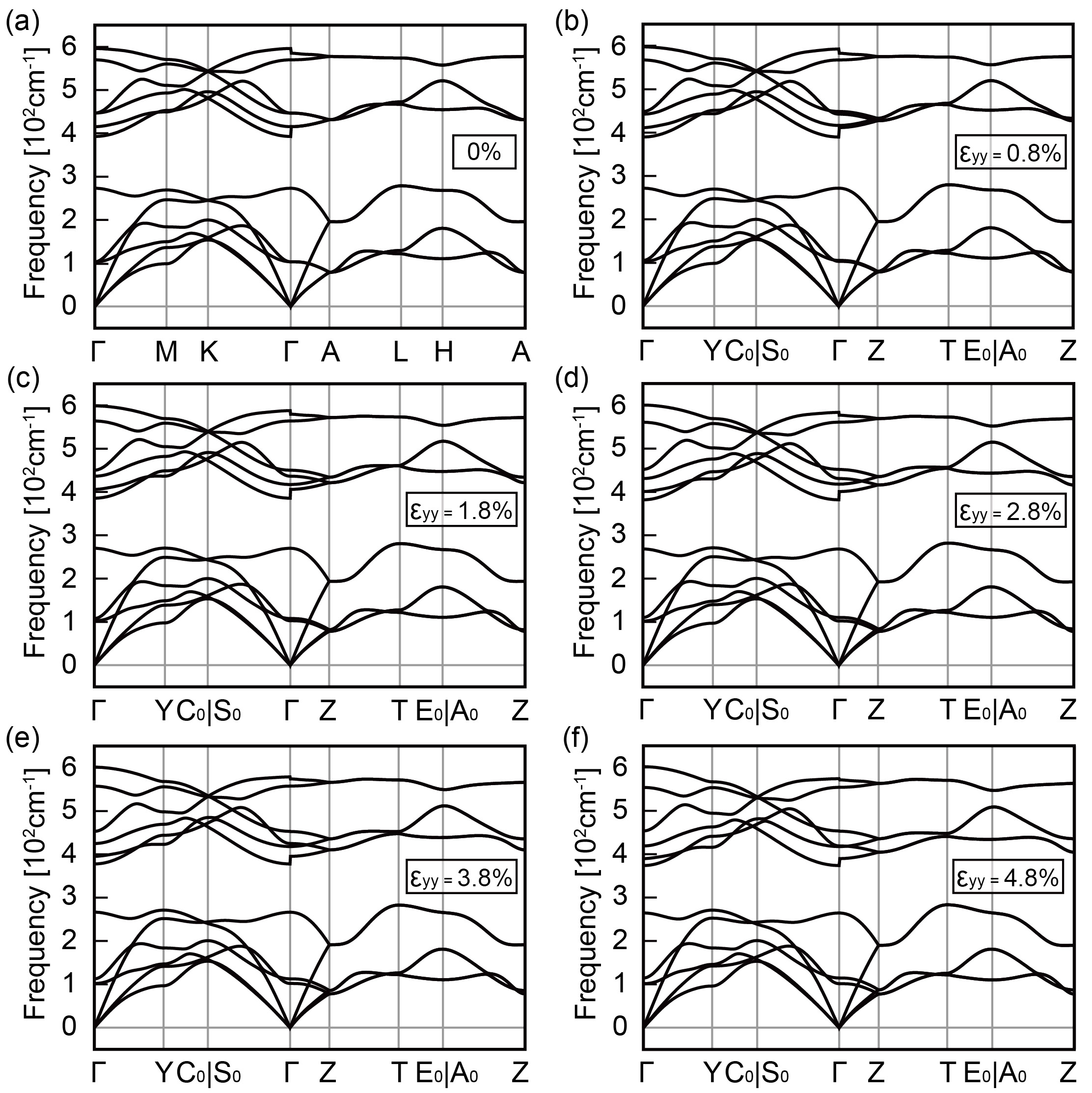}
\caption{\label{fig:A2}
Phonon dispersions of ZnO along high-symmetry lines. (a) Pristine ZnO in the hexagonal Brillouin zone. (b)--(f) $\varepsilon_{yy}$-strained ZnO in the orthorhombic Brillouin zone; the applied strain is indicated in each panel.
}
\end{figure}

We compute the electronic and phonon band structures of $\varepsilon_{yy}$-strained ZnO along high-symmetry lines in the Brillouin zone. Figure \ref{fig:A1} (a) shows the electronic band structure of the unstrained hexagonal cell along the path $\Gamma$-$M$-$K$-$\Gamma$-$A$-$L$-$H$-$Z$. Figure\,\ref{fig:A1} (b)--(f) show the corresponding band structures for the strained orthorhombic cell along the symmetry-equivalent path $\Gamma$-$Y$-$C_0|S_0$-$\Gamma$-$Z$-$T$-$E_0|A_0$-$Z$. Figure \ref{fig:A2} presents the phonon dispersions along the same high-symmetry paths as in Fig.\,\ref{fig:A1}. Overall, both the electronic and phonon band structures evolve smoothly with increasing strain, with no abrupt or discontinuous changes over the strain range considered.

\section{Details of Wannier interpolation}

\begin{table}[b]
  \caption{Frozen energy window from $E_\text{min}$ to $E_\text{max}$ for each $\varepsilon_{yy}$ strain condition.}
  \label{tab:wan}
  \begin{ruledtabular}
  \centering 
  \begin{tabular}{ccc}
    $\varepsilon_{yy}$ Strain (\%) &  $E_\text{min}$($\mathrm{eV}$) &
    $E_\text{max}$($\mathrm{eV}$)
    \\
    \hline
     0.0  &  $-3.76$ &$11.47$ \\
     0.8  &  $-3.86$ &$11.37$ \\
     1.8  &  $-3.76$ &$11.47$ \\
     2.8  &  $-3.86$ &$11.47$ \\
     3.8  &  $-3.89$ &$11.44$ \\
     4.8  &  $-3.86$ &$11.47$ \\
  \end{tabular}
  \end{ruledtabular}
\end{table}

After testing multiple setups, we identified frozen energy windows and initial projectors that yield reliable Wannier Hamiltonians for each strain condition. In total, 14 hybrid orbitals were used as initial projectors: 
for one Zn site,
four $sp^3d^2$ type hybrid projectors of $|\varphi^n_\text{Zn}\rangle$ where $n=2,3,4,5$ (the number $n$ follows the convention of {\sc{Wannier90}} code), 
$
|\varphi^2_\text{Zn}\rangle=
\frac{1}{\sqrt{6}}
|s\rangle
+
\frac{1}{\sqrt{2}}
|p_x\rangle
-
\frac{1}{\sqrt{12}}
|d_{z^2}\rangle
+
\frac{1}{2}
|d_{x^2-y^2}\rangle
$,
$|\varphi^3_\text{Zn}\rangle=
\frac{1}{\sqrt{6}}
|s\rangle
-
\frac{1}{\sqrt{2}}
|p_y\rangle
-
\frac{1}{\sqrt{12}}
|d_{z^2}\rangle
-
\frac{1}{2}
|d_{x^2-y^2}\rangle
$,
$|\varphi^4_\text{Zn}\rangle=
\frac{1}{\sqrt{6}}
|s\rangle
+
\frac{1}{\sqrt{2}}
|p_y\rangle
-
\frac{1}{\sqrt{12}}
|d_{z^2}\rangle
-
\frac{1}{2}
|d_{x^2-y^2}\rangle
$,
$|\varphi^5_\text{Zn}\rangle=
\frac{1}{\sqrt{6}}
|s\rangle
-
\frac{1}{\sqrt{2}}
|p_z\rangle
+
\frac{1}{\sqrt{3}}
|d_{z^2}\rangle
$.
For an oxygen site, five $sp^3d^2$ type hybrid projectors of $|\varphi^n_\text{O}\rangle$ ($n=1,2,3,5,6)$, 
$
|\varphi^1_\text{O}\rangle=
\frac{1}{\sqrt{6}}
|s\rangle
-
\frac{1}{\sqrt{2}}
|p_x\rangle
-
\frac{1}{\sqrt{12}}
|d_{z^2}\rangle
+
\frac{1}{2}
|d_{x^2-y^2}\rangle
$,
$
|\varphi^2_\text{O}\rangle=
\frac{1}{\sqrt{6}}
|s\rangle
+
\frac{1}{\sqrt{2}}
|p_x\rangle
-
\frac{1}{\sqrt{12}}
|d_{z^2}\rangle
+
\frac{1}{2}
|d_{x^2-y^2}\rangle
$,
$|\varphi^3_\text{O}\rangle=
\frac{1}{\sqrt{6}}
|s\rangle
-
\frac{1}{\sqrt{2}}
|p_y\rangle
-
\frac{1}{\sqrt{12}}
|d_{z^2}\rangle
-
\frac{1}{2}
|d_{x^2-y^2}\rangle
$,
$|\varphi^5_\text{Zn}\rangle=
\frac{1}{\sqrt{6}}
|s\rangle
+
\frac{1}{\sqrt{2}}
|p_y\rangle
-
\frac{1}{\sqrt{12}}
|d_{z^2}\rangle
-
\frac{1}{2}
|d_{x^2-y^2}\rangle
$,
$|\varphi^6_\text{Zn}\rangle=
\frac{1}{\sqrt{6}}
|s\rangle
-
\frac{1}{\sqrt{2}}
|p_z\rangle
+
\frac{1}{\sqrt{3}}
|d_{z^2}\rangle
$.
For the other oxygen site, five $sp^3d$ type hybrid projectors of $|\phi^n_\text{O}\rangle$ ($n=1,2,3,4,5)$, 
$|\phi^1_{O}\rangle 
=\frac{1}{\sqrt{3}}|s\rangle
-
\frac{1}{\sqrt{6}}|p_x\rangle
+
\frac{1}{\sqrt{2}}|p_y\rangle
$,
$|\phi^2_{O}\rangle 
=\frac{1}{\sqrt{3}}|s\rangle
-
\frac{1}{\sqrt{6}}|p_x\rangle
-
\frac{1}{\sqrt{2}}|p_y\rangle
$,
$|\phi^3_{O}\rangle 
=\frac{1}{\sqrt{3}}|s\rangle
+
\frac{2}{\sqrt{6}}|p_x\rangle
$,
$|\phi^4_{O}\rangle 
=\frac{1}{\sqrt{2}}|p_z\rangle
+
\frac{2}{\sqrt{2}}|d_{z^2}\rangle
$,
$|\phi^5_{O}\rangle 
=-\frac{1}{\sqrt{2}}|p_z\rangle
+
\frac{1}{\sqrt{2}}|d_{z^2}\rangle
$.

\begin{figure}[t]
\centering
\includegraphics[width=0.48\textwidth]{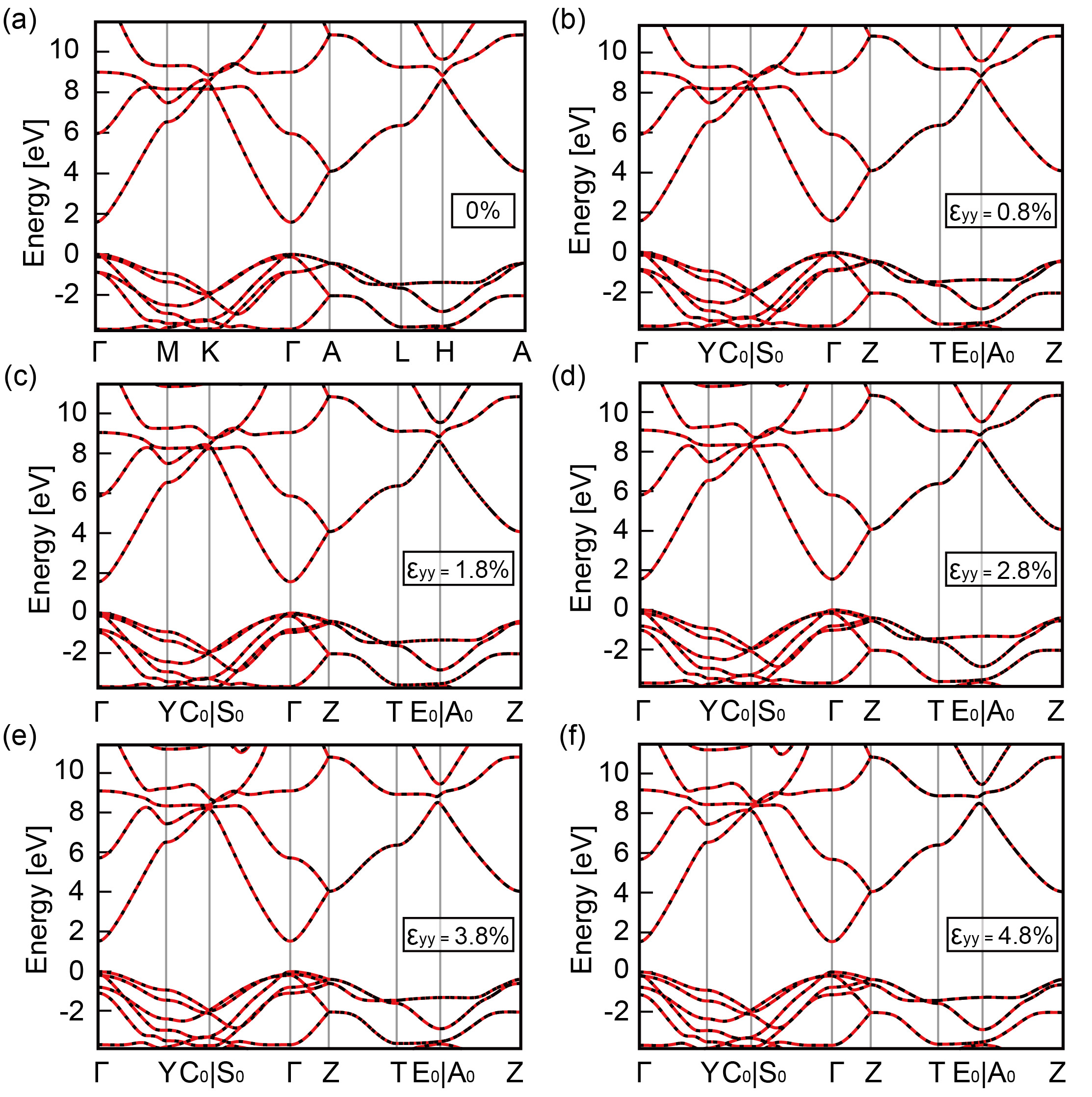}
\caption{\label{fig:A3}
Comparison of Wannier-interpolated and DFT band structures along high-symmetry lines. Red solid lines show the Wannier bands, and black dotted lines show the corresponding DFT bands. (a) Pristine ZnO. (b)--(f) $\varepsilon_{yy}$-strained ZnO, with the applied strain indicated in each panel.
}
\end{figure}

The frozen-window parameters for each strained structure are listed in Table~\ref{tab:wan}, where the Fermi level is set to the valence-band maximum. The Wannierization converged within $10^4$ iterations. Figure~\ref{fig:A3} compares the Wannier-interpolated bands with the underlying DFT bands. The Wannier bands (red solid lines) closely reproduce the DFT dispersion (black dashed lines), with particularly good agreement near the band edges around $\Gamma$, where the dominant contributions to transport and the near-edge optical response originate.

\section{Effective mass along the three principle axis}
\begin{figure}[t]
\centering
\includegraphics[width=0.48\textwidth]{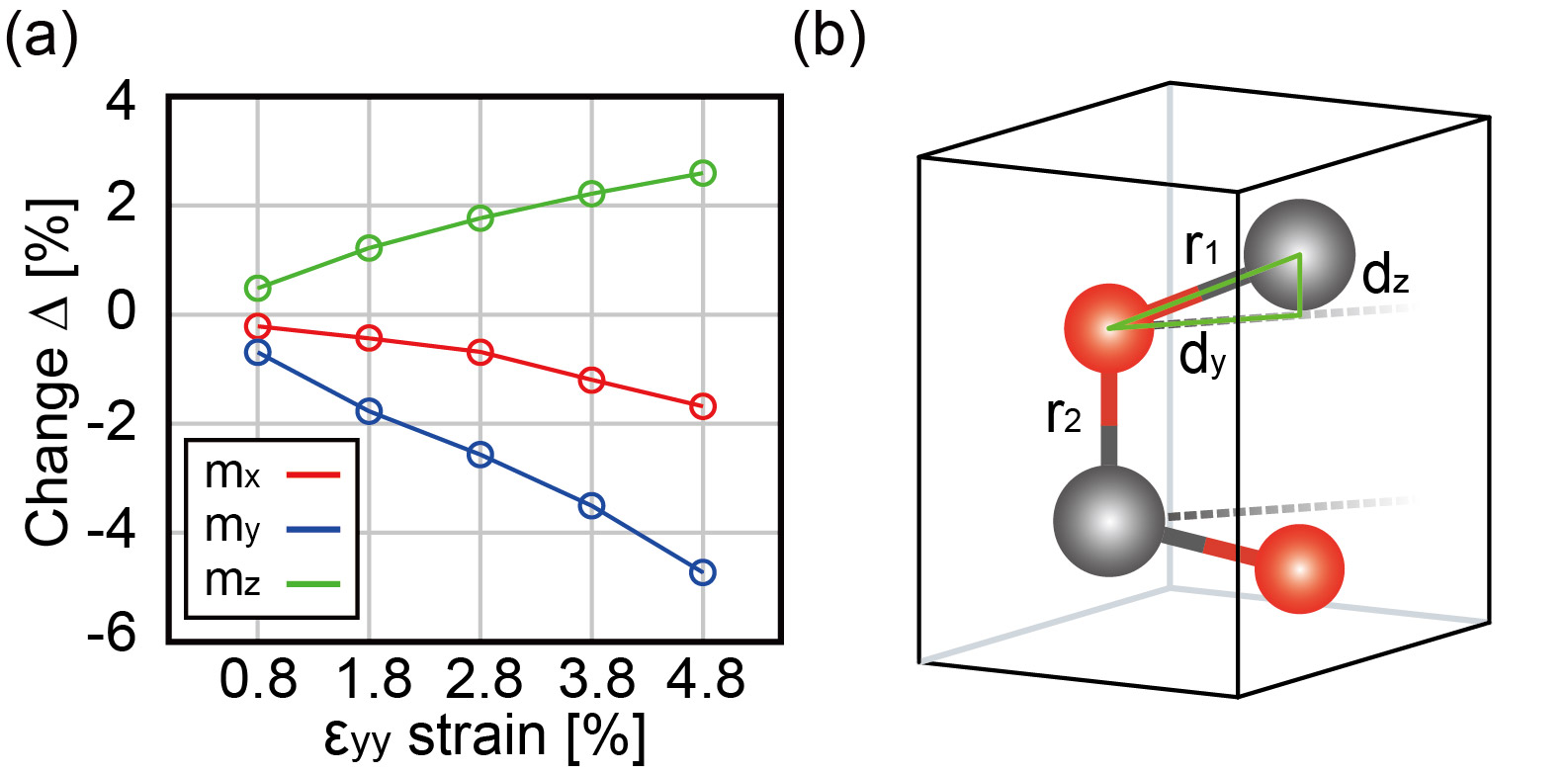}
\caption{\label{fig:A4}
Strain dependence of the conduction-band effective mass and the associated structural changes. (a) Relative change in the effective mass extracted from the inverse band curvature along the $x$ (red), $y$ (blue), and $z$ (green) directions, under $\varepsilon_{yy}$ strain. (b) Schematic illustration of the Zn--O bonds under $\varepsilon_{yy}$ strain.
}
\end{figure}

We examine the $\varepsilon_{yy}$ strain dependence of the electron effective mass along the three principal axes. The effective mass is extracted from the inverse curvature of the lowest conduction band dispersion along each direction and is normalized to the pristine ZnO value for comparison. Figure~\ref{fig:A4} (a) summarizes the relative changes in the effective masses along $x$ (red), $y$ (blue), and $z$ (green). Under tensile $\varepsilon_{yy}$ strain, both in-plane components decrease, with a larger reduction along the strain direction, $m_y$, than along the perpendicular in-plane direction, $m_x$. By contrast, the out-of-plane effective mass $m_z$ increases. The directional bandwidth shows no substantial reduction along the strain direction, indicating that the effective hopping amplitude is nearly preserved. This behavior can be understood from the directional dependence of the hopping matrix element in the Slater--Koster form \cite{slater1954simplified}. As shown in Fig.~\ref{fig:A4} (b), the tilted Zn--O $sp$-$\sigma$ bond $r_1$ lies in the $yz$ plane, with projected distances $d_y$ and $d_z$. The corresponding hopping contribution is proportional to $(d_y/r_1)t_{sp\sigma}$ along the $y$ direction and to $(d_z/r_1)t_{sp\sigma}$ along the $z$ direction, where $t_{sp\sigma}$ is the bond integral. Under tensile $\epsilon_{yy}$ strain, the increase in $d_y$ compensates for the elongation of $r_1$, preventing a significant reduction of the effective hopping along the strain direction. Therefore, the strain-induced changes in the effective mass can be understood as the combined result of modified lattice periodicity and reduced band gap, with the effective hopping amplitude remaining nearly unchanged.

\section{Strain-direction dependence of ZnO}
\begin{figure}[t]
\centering
\includegraphics[width=0.48\textwidth]{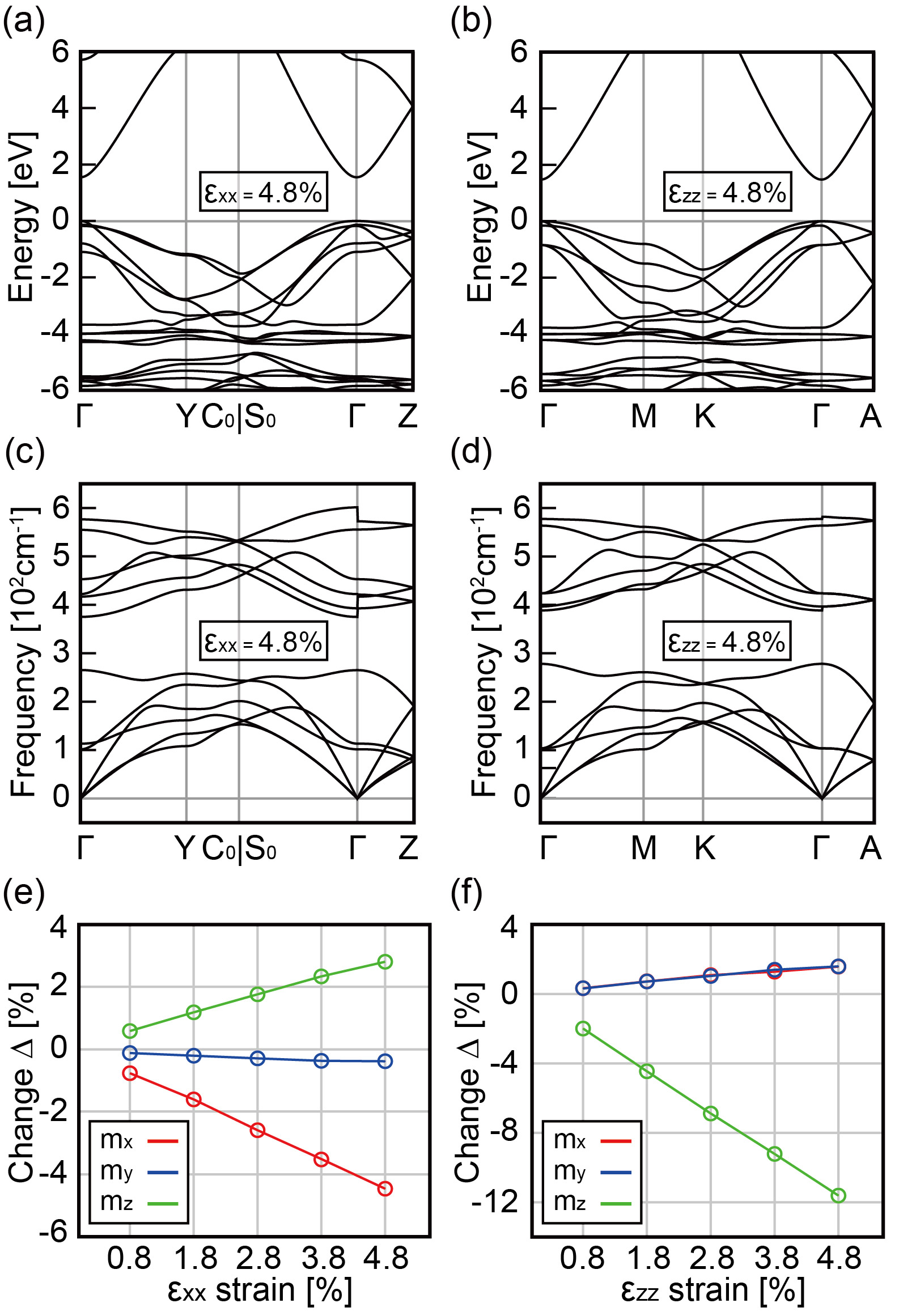}
\caption{\label{fig:A5}
Electronic and phonon band structures of ZnO under tensile strain. The left panels, (a), (c), and (e), correspond to $\varepsilon_{xx}$ strain, while the right panels, (b), (d), and (f), correspond to $\varepsilon_{zz}$ strain. (a),(b) Electronic band structures. (c),(d) phonon band structures. (e),(f) Relative change in the effective mass extracted from the inverse band curvature along the $x$ (red), $y$ (blue), and $z$ (green) directions. 
}
\end{figure}

\begin{table}[t]
\caption{\label{tab:xz_strained_cell} Relaxed lattice parameters and self-consistent Hubbard $U_p$ values for ZnO under tensile strain. The upper and lower blocks correspond to $\varepsilon_{xx}$ and $\varepsilon_{zz}$ strain, respectively. Here, $a'$ ($c'$) denote the in-plane (out-of-plane) lattice constants, and $l$ is the in-plane distortion parameter defined in the main text.}
\begin{ruledtabular}
\begin{tabular}{c|cccc}
Strain (\%) & $a'$ (\text{\AA}) & $l$ & $c'$ (\text{\AA}) & $U_p$ (eV) \\
\hline
0.8  & 3.20 & 1.71 & 5.09 & 2.8816 \\
1.8  & 3.23 & 1.69 & 5.07 & 2.8823 \\
2.8  & 3.27 & 1.66 & 5.06 & 2.8831 \\
3.8  & 3.30 & 1.64 & 5.04 & 2.8846 \\
4.8  & 3.33 & 1.61 & 5.03 & 2.8868 \\
\hline\hline
0.8  & 3.17 & $\sqrt{3}$ & 5.14 & 2.8835 \\
1.8  & 3.16 & $\sqrt{3}$ & 5.19 & 2.8872 \\
2.8  & 3.15 & $\sqrt{3}$ & 5.24 & 2.8917 \\
3.8  & 3.14 & $\sqrt{3}$ & 5.29 & 2.8970 \\
4.8  & 3.13 & $\sqrt{3}$ & 5.34 & 2.9030 \\
\end{tabular}
\end{ruledtabular}
\end{table}

\begin{figure}[t]
\centering
\includegraphics[width=0.48\textwidth]{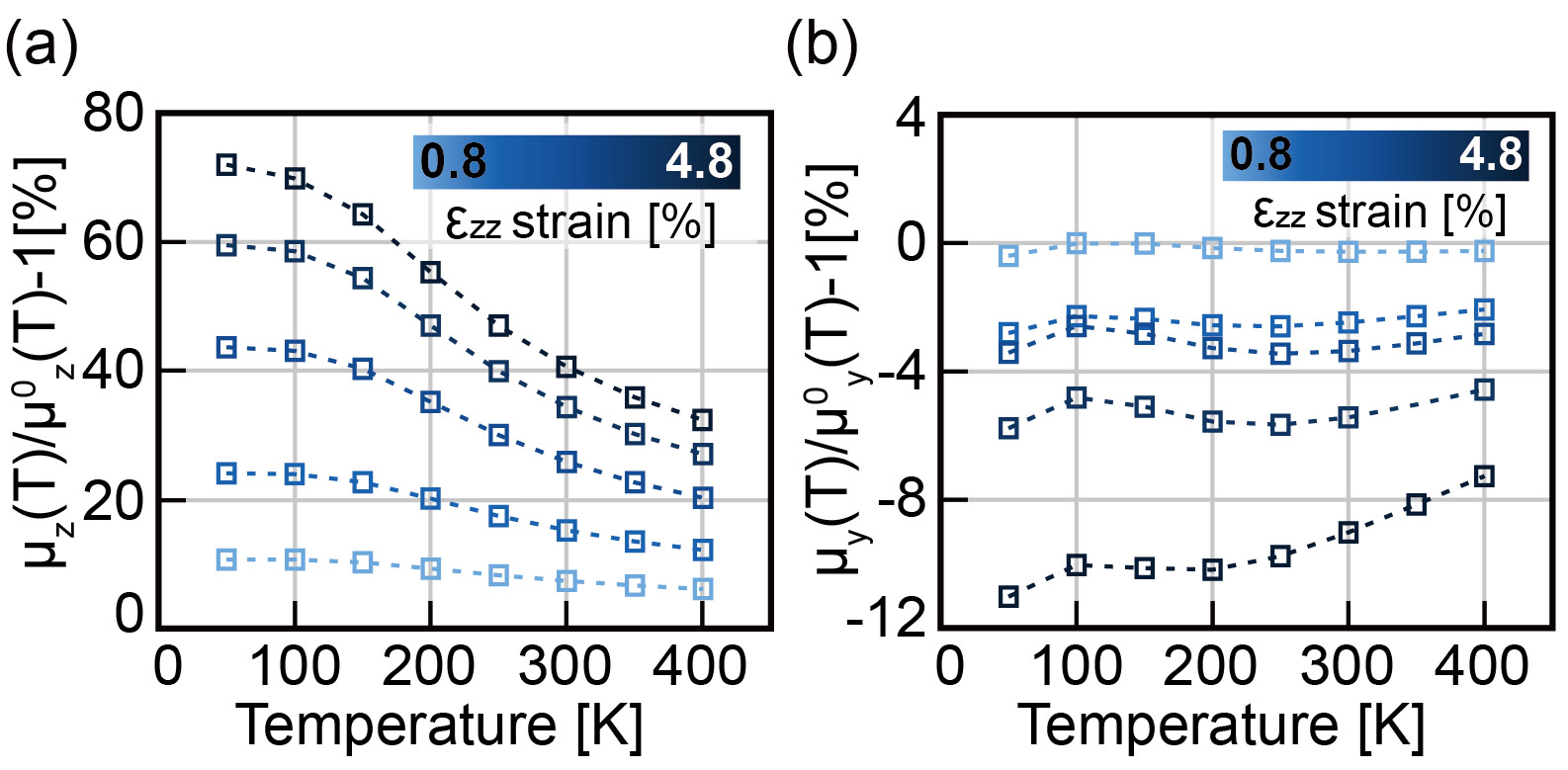}
\caption{\label{fig:A6} Transport properties of ZnO under $\varepsilon_{zz}$ strain. (a),(b) Relative change of the phonon-limited electron mobility along the (a) $z$ and (b) $y$ directions under $\varepsilon_{zz}$ strain. The changes in percent are shown with respect to $\mu_i^{0}(\mathrm{T})$, which denotes the pristine mobility at each temperature. The strained mobilities are colored in shades of blue, with darker colors indicating larger $\varepsilon_{zz}$ strain ranging from $0.8\%$ to $4.8\%$.
}
\end{figure}

We examine the structural, electronic, and vibrational properties of ZnO under $\varepsilon_{xx}$ and $\varepsilon_{zz}$ strain. Table~\ref{tab:xz_strained_cell} summarizes the relaxed lattice parameters and the corresponding self-consistent Hubbard parameters $U_p$. For $\varepsilon_{xx}$ strain, the angle $\theta$ between the in-plane lattice vectors decreases, which is reflected in the reduction of the in-plane distortion parameter $l$. The lattice parameter along the strain direction increases, while the lattice parameters along the two perpendicular directions decrease due to the Poisson effect. We find that $U_p$ remains robust against strain, consistent with the $\varepsilon_{yy}$-strained case discussed in the main text. For $\varepsilon_{zz}$ strain, the hexagonal symmetry is preserved, and therefore $l$ is fixed to be $\sqrt{3}$. In this case, the out-of-plane lattice parameter $c'$ increases, while the in-plane lattice parameter slightly decreases.

Figure~\ref{fig:A5} shows the electronic and phonon band structures of strained ZnO. Panels (a) and (b) show the electronic band structures at the maximum strain considered for the $\varepsilon_{xx}$ and $\varepsilon_{zz}$, respectively. The electronic structure of the $\varepsilon_{xx}$-strained case resembles that of the $\varepsilon_{yy}$-strained case in the main text, apart from the different orientation of the orthorhombic cell. In contrast, $\varepsilon_{zz}$ strain preserves the symmetry of pristine ZnO, while reducing the band gap. The phonon band structures shown in Fig.~\ref{fig:A5} (c) and (d) confirm that the strained structures are dynamically stable.

We also calculate the effective mass of the conduction electron from the inverse band curvature. As shown in Fig.~\ref{fig:A5} (e), $\varepsilon_{xx}$ strain reduces the effective mass primarily along the $x$ direction, whereas the $\varepsilon_{yy}$ strain mainly reduces the effective mass along the $y$ direction. For both in-plane strain directions, the out-of-plane effective mass $m_z$ increases. Under $\varepsilon_{zz}$ strain, shown in Fig.~\ref{fig:A5} (f), the effective masses along the $x$ and $y$ directions are nearly identical because the rotational symmetry is preserved near the $\Gamma$ point. In this case, $m_z$ decrease substantially, while the in-plane effective masses $m_x$ and $m_y$ increase. The bandwidth along the $z$ direction shows that the effective hopping is slightly reduced under $\epsilon_{zz}$ strain. Although the increase in the projected distance $d_z$ partly compensates for the elongation of the $r_1$ bond in the hopping contribution $(d_z/r_1)t_{sp\sigma}$, the $z$-aligned bond $r_2$ is also elongated and reduces the hopping along the strain direction. This suggests that the substantial decrease in $m_z$ is unlikely to originate from enhanced hopping. It is instead attributed to the increased out-of-plane lattice periodicity and the $7\%$ reduction of the band gap, which strengthens interband coupling.

The phonon-limited mobility of ZnO under $\varepsilon_{zz}$ strain is shown in Fig.~\ref{fig:A6}. As expected from the substantial reductions in the band gap and $m_z$, the mobility in the $z$ direction increases with strain, reaching a $40.6\%$ enhancement at 300~K for the maximum strain. This is about twice the increase in the mobility component along the strain direction under $\varepsilon_{yy}$ strain shown in Fig.~\ref{fig:mob} (b), while the two cases show similar strain and temperature dependences. The in-plane mobility $\mu_y$ under $\varepsilon_{zz}$ strain decreases, similar to the out-of-plane mobility $\mu_z$ under $\varepsilon_{yy}$ strain shown in Fig.~\ref{fig:mob} (c). Overall, $\varepsilon_{zz}$ strain provides an out-of-plane counterpart to the $\varepsilon_{yy}$ strain studied in the main text.

\bibliography{./refs_rev}

\end{document}